\documentclass[12pt]{article}
\usepackage{graphicx}
\usepackage{amsmath,amsthm,amscd,amssymb}

\begin{document}

\title{Crowdsourcing, Attention and Productivity}

\author{Bernardo A. Huberman, Daniel M. Romero and Fang Wu\\ \small{Social Computing Lab, HP Laboratories, Palo Alto, CA 94304}}

\maketitle

\bigskip
\begin{abstract}

The tragedy of the digital commons does not prevent the copious voluntary production of content that one witnesses in the web. We show through an analysis of a massive data set from \texttt{YouTube} that the productivity exhibited in crowdsourcing exhibits a strong positive dependence on attention, measured by the number of downloads. Conversely, a lack of attention leads to a decrease in the number of videos uploaded and the consequent drop in productivity, which in many cases asymptotes to no uploads whatsoever.  Moreover, uploaders compare themselves to others when having low productivity and to themselves when exceeding a threshold.
\end{abstract}

\pagebreak
We are witnessing an inversion of the traditional way by which content has been generated and consumed over the centuries. From photography to news and encyclopedic knowledge, the centuries-old pattern has been one in which a relatively few people and organizations produce content and most people consume it. With the advent of the web and the ease with which one can migrate content to it, that pattern has reversed, leading to a situation whereby millions create content in the form of blogs, news, videos, music, etc.~and relatively few can attend to it all. This phenomenon, which goes under the name of \emph{crowdsourcing}, is exemplified by websites such as \texttt{Digg}, \texttt{Flicker}, \texttt{YouTube}, and \texttt{Wikipedia}, where content creation without the traditional quality filters manages to produce sought out movies, news and even knowledge that rivals the best encyclopedias. That such content is valued is confirmed by the fact that access to these sites accounts for a sizable percentage of internet traffic. For example, as of June, 2007 \texttt{YouTube} alone comprised approximately 20\% of all HTTP traffic, or nearly 10\% of all traffic on the Internet \cite{anderson}.

What makes crowdsourcing both interesting and puzzling is the underlying dilemma facing every contributor, which is best exemplified by the well-known \emph{tragedy of the commons}. In such dilemmas, a group of people attempts to provide a common good in the absence of a central authority. In the case of crowdsourcing, the common good is in the form or videos, music, or encyclopedic knowledge that can be freely accessed by anyone. Furthermore, the good has jointness of supply, which means that its consumption by others does not affect the amounts that other users can use.
And since it is nearly impossible to exclude non contributors from using the common good, it is rational for individuals not to upload content and free ride on the production of others. The dilemma ensues when every individual can reason this way and free ride on the efforts of others, making everyone worse off---thus the tragedy of the digital commons \cite{gnutella,asvanund,hughes,glance,levine}.

And yet paradoxically, there is ample evidence that while the ratio of contributions to downloads is indeed small, the growth in content provision persists at levels that are hard to understand if analyzed from a public goods point of view. One possible explanation for this puzzling behavior, which we explore in this paper, is that those contributing to the digital commons perceive it as a private good, in which payment for their efforts is in the form of the attention that their content gathers in the form of media downloads or news clicked on. As it has been shown, attention is such a valued resource that people are often willing to forsake financial gain to obtain it \cite{status}. In the world of academia, for example, attention is often its main currency, for we publish to get the attention of others, we cite so that other researcher's work get attention, and we cherish the prominence of great work if only because of the attention it gathers \cite{franck}. Similarly, within online communities, status and recognition have been shown to be very important motivators for contributing \cite{lampel}.\footnote{Another important instance is open source software development.  Several studies have shown however, that open source projects are characterized by a very small core of contributors \cite{mockus} where the free-riding problem is not acute.}

If attention is indeed the main driver of contributions to the digital commons, one should be able to observe a correlation between the rate at which content is generated and the number of downloads. And if in addition a causal relation between the two does exist, we expect that those contributors that have a high level of downloads will continue to contribute, whereas those who see a decline in the attention that their content is receiving will decrease their productivity.

In order to investigate this conjecture we collected data from \texttt{YouTube}, a popular website that allows its users to upload, view, and share video clips. After a \texttt{YouTube} user uploads a video, a ``view count'' number is immediately displayed next to the video title, which measures how many times it has been watched. Our dataset contained 9,896,816 videos submitted by 579,471 users by April 30, 2008. For each video upload we obtained its datestamp, the uploader's id, and the final view count.

To study the dynamic interplay between productivity and attention, we partitioned time into 2-week periods, starting when they upload their first video and ending when they upload their last one. A common pattern we observed is that most periods between a contributor's first and last uploads contain no uploads at all (on average, 66\% of these periods are empty), indicating an intermittent productivity. Because of the bursty nature of our data, we considered only the ``active'' periods for each contributor (i.e.~periods containing at least one upload), and labeled them as $t=1,2,\dots$.

We measured the productivity of each contributor by the number of videos $n_t$ she uploads during the $t$'th active period, and the attention she receives by the average number of views $v_t$ of the $n_t$ videos. In other words we wanted to establish how $v_t$ affects $n_{t+1}, n_{t+2}, \dots$, which provides dynamical information on how each contributor responds to different amounts of attention.


We  conducted a robust linear regression  $\{n_{t+1}\}_{t=1}^T \sim \alpha \{\log_{10}v_t\}_{t=1}^T + \beta$ for each contributor that was active for $T > 10$ periods \cite{robust}. (Because the view counts varied over many orders of magnitudes, it made sense to consider $\log_{10}v_t$ instead of $v_t$.) We thus collected
76,462 $\alpha$ values and conducted a $t$-test of the null hypothesis that the $\alpha$ values come from a normal distribution with non-positive mean. The resulting $p$-value is less than 0.001, suggesting that the null hypothesis can be rejected. We also conducted the same test with different choices of $T$, and observed that as long as $T>10$ the $p$-value was always less than 0.001. Hence, for those contributors who were active for a minimum number of periods, the more views they received in one period, the more videos they uploaded during the following period.

A more direct approach to test our conjecture is to measure the change in each contributor's productivity at different attention levels. For each contributor who was active for at least two different periods, define $\bar v= \mbox{median} \{v_t\}_{t=1}^{T-1}$ as her median received attention, where $T$ is her number of active periods. According to this definition, all periods can be divided into two groups of equal size, $\lfloor (T-1)/2 \rfloor$: the ``good periods'' in which she receives higher than usual attention ($G = \{s: v_s>\bar v\}$), and the ``bad periods'' in which she receives lower than usual attention ($B=
\{s: v_s<\bar v\}$).

Let $n^G= \frac 1{\lfloor (T-1)/2 \rfloor} \sum_{s\in G} n_{s+1}$ denote the average productivity following a good period, and let $n^B= \frac 1{\lfloor (T-1)/2 \rfloor} \sum_{s\in B} n_{s+1}$ denote the average productivity following a bad period. With these definitions the difference $\Delta = n^G - n^B$  measures the change of a contributor's productivity between different attention levels. If $\Delta>0$ contributors upload more videos after obtaining more views, and if $\Delta<0$ the opposite is true.

Figure \ref{hist_delta} shows the histogram of the different 20,061 $\Delta$ values for the group of contributors who were active for 2 to 9 periods. A $t$-test of the null hypothesis that $\Delta\le 0$ yields a $p$-value less than $0.001$, leading to rejection of the null hypothesis. Thus on average each contributor becomes more productive after a good period than a bad period.

{\begin{figure}[htb]
\begin{center}
\includegraphics[width = 1 \textwidth, height =.6 \textwidth] {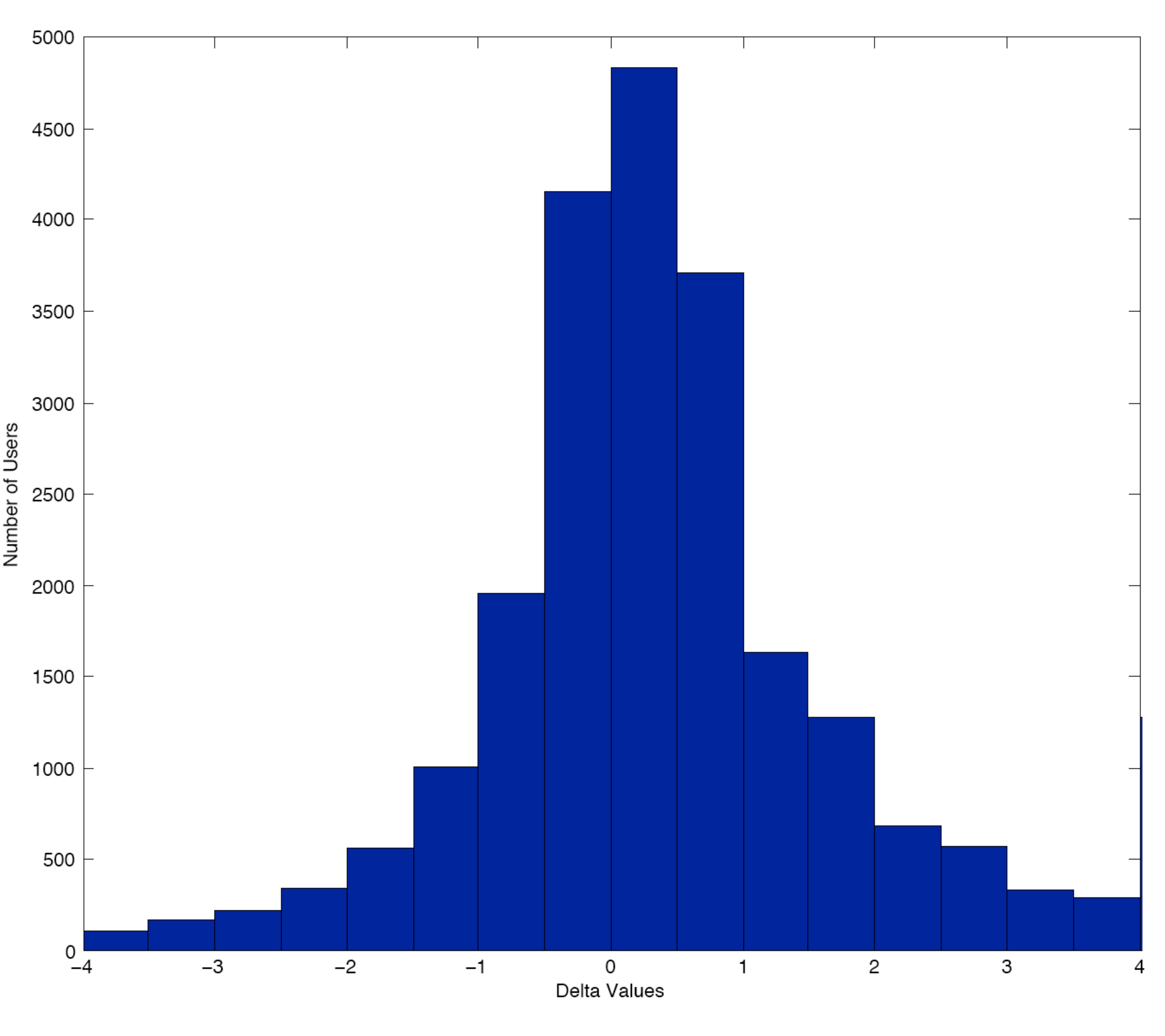} \end{center} \caption{\small{\textbf{Histogram of contributor's $\Delta$ values for contributors that were active from 2 to 9 weeks.
Notice that the maximum of the histogram is shifted to the right of the origin.  The null hypothesis that data comes form a normal distribution with non-positive mean, can be rejected with $p$-value less than 0.001.}}} \label{hist_delta} \end{figure}

Figure \ref{hist_delta} indicates that each contributor tends to become more productive after receiving a number of views that exceeds her own normal performance. One can also test whether his productivity increases as she outperforms the average contributor in the general population. To do so, we measured the average view count of all videos in our dataset, which is given by $\bar v = 10000$, and used it to measure the productivity difference between good periods (more than 10000 views on average) and bad periods (less than 10000 views on
average) through the quantity $\Delta=n^G-n^B$. We divided the contributors into several different groups depending on their number of active periods, and tested the null hypothesis ``$\Delta \le 0$'' for each group. Table \ref{constant threshold} shows the results from these tests, including the number of contributors considered in each subgroup, the mean of the $\Delta$ values, and the $p$-values of the null hypothesis. Notice that the $p$-values are very small for most groups, which supports our hypothesis that a competitive factor enters into the productivity of contributors. Also note in Table \ref{constant threshold} that the mean of $\Delta$ \emph{decreases} as the number of active weeks increases, indicating that those people who made relatively few contributions care more about their relative performance against other contributors.

For comparison purposes we also tested the same null hypothesis for $\bar v = \mbox{median} \{v_t\}_1^{T-1}$ (i.e.~the median view count of each contributor) which is not constant but varies from contributor to contributor. The results are listed in Table \ref{varying threshold}. We see that in this case the mean of $\Delta$ \emph{increases} as the number of active weeks increases, indicating that the productive ones care more about how they have improved their own performance, rather than comparing with the rest of the community.

\begin{table}[ht]
\centering
\begin{tabular}{c c c c}
\hline\hline
number of active weeks & number of contributors & $\Delta$-mean & $p$-value\\ [0.5ex] \hline
2-9 & 20061 & .59 & $< .001$ \\
10-19 & 24517 & .58 & $< .001$ \\
20-29 & 7789 & .32 & $< .001$ \\
30-39 & 2153 & .09 & .11 \\
40-70 & 515 & -.05 & .61 \\[1ex]
\hline
\end{tabular}
\caption{\small{\textbf{Tests of the null hypothesis ``$\Delta \le 0$'', where $\Delta=n^G-n^B$ measures the productivity difference between a contributor's good periods (in which her contributions received more than 10000 views on average) and bad periods (less than 10000 views on average). As the number of active weeks increases, the mean of $\Delta$ decreases.}}} \label{constant threshold} \end{table}

\begin{table}[ht]
\centering
\begin{tabular}{c c c c}
\hline\hline
number of active weeks & number of contributors & $\Delta-$mean & $p$-value\\ [0.5ex] \hline
2-9 & 85949 & .01 & $.14$ \\
10-19 & 68317 & .15 & $< .001$ \\
20-29 & 14757 & .18 & $< .001$ \\
30-39 & 3303 & .20 & $< .001$ \\
40-70 & 673 & .26 & $< .01$ \\[1ex] \hline \end{tabular} \caption{\small{\textbf{Tests of the null hypothesis ``$\Delta \le 0$'', where $\Delta=n^G-n^B$ measures the productivity difference between a contributor's good periods (in which her contributions received more than her median view count) and bad periods (less than her median view count).
As the number of active weeks increases the mean of $\Delta$ increases.}}} \label{varying threshold} \end{table}

While the observed correlations between attention and productivity suggest a trend, they do not imply a causal relation between them. In fact, it is not clear whether an increase in attention causes productivity as a whole to grow or vice-versa. In order to clarify this issue we used a Granger causality test, which is a statistical tool that determines causality in terms of prediction accuracy \cite{granger}. Given two signals $X_1$ and $X_2$, we say that $X_1$ G-causes $X_2$ if past values of $X_1$ contain information that helps predict future values of $X_2$.  It is important to note that Granger causality is only meaningful if only found in one direction, i.e.~$X_1$ G-causes $X_2$ but $X_2$ does not G-cause $X_1$. If on the other hand Granger causality is found in both directions it is likely that $X_1$ and $X_2$ are only correlated and that the correlation is caused by a third signal.

In order to determine the causal relation between attention and productivity, we defined $\bar v_t$ to be the average of the all contributor's views during their $t$'th active period, and similarly we let $\bar n_t$ be the average of all contributor's videos uploads during their $t$'th active week. We then conducted a Granger causality test of the hypothesis that $\bar v_t$ G-causes $\bar n_t$, which resulted in a $p$-value of 0.01, and of the hypothesis that $\bar n_t$ G-causes $\bar v_t$, which gave a $p$-value of 0.61. This result shows that attention plays a determinant role in the productivity of those uploading videos.


{\begin{figure}[ht]
\begin{center}
\includegraphics[width = 1 \textwidth, height =.6 \textwidth] {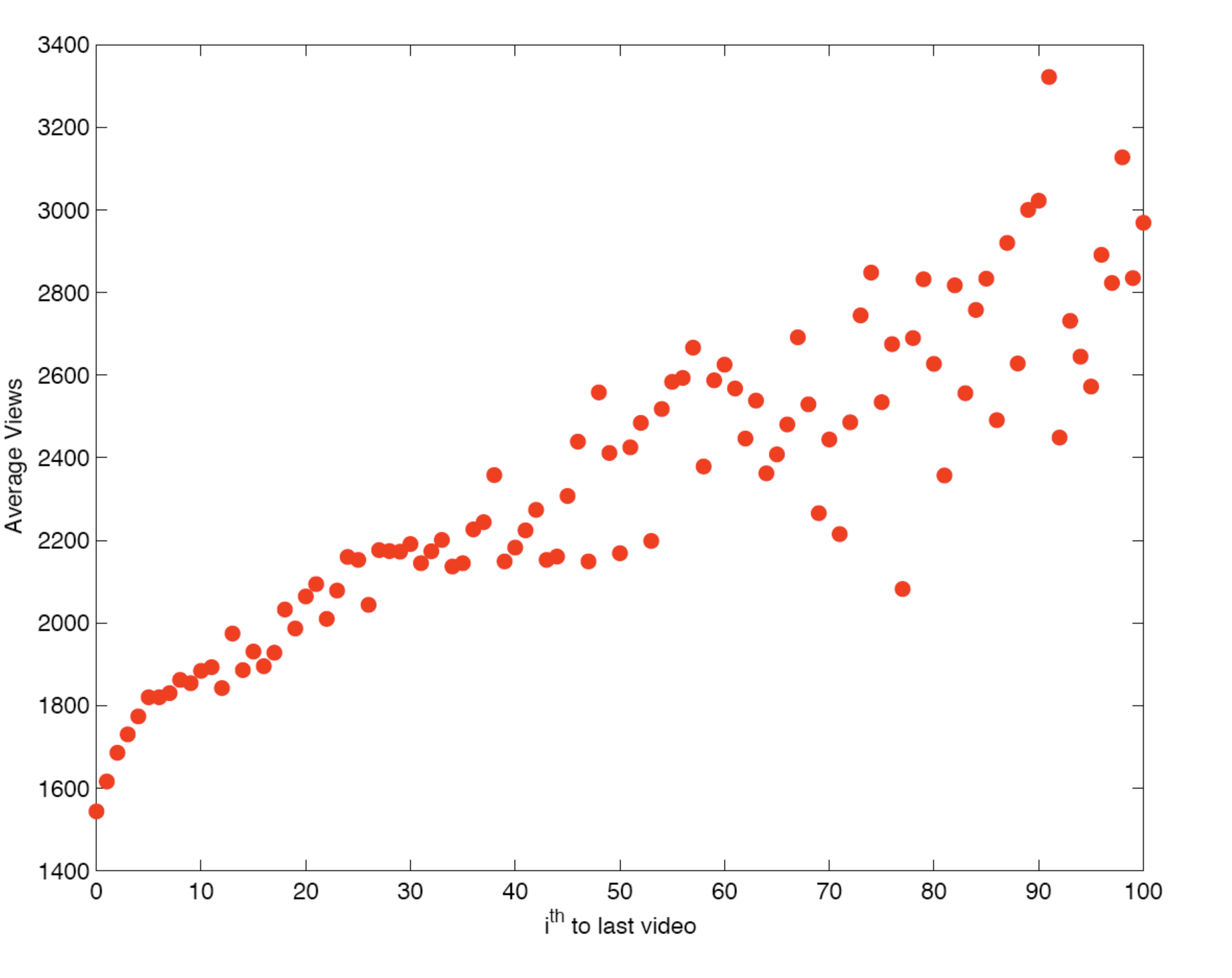}
\end{center}
\caption{\small{\textbf{Average number of views vs.~$i$'th to last video. The origin represents the last video. The average number of views decreases linearly as contributers approach their last video with correlation of 0.90.}}} \label{nth_video_views_n_Jan_1_2008_vmax_10000}
\end{figure}

Finally, since it is a common observation that many contributors stop uploading videos, we decided to test if this behavior was due to the small number of downloads their videos receive. To do so we considered all the contributors in our dataset that had not uploaded any videos during the four months previous to the date the data was collected.

Figure \ref{nth_video_views_n_Jan_1_2008_vmax_10000} shows the number of average views as a function of the $i$'th to last video. As can be seen, as contributors approach their last video upload at the origin, the average number of previous views of their videos exhibited a marked linear decrease. This confirms our conjecture that decreasing attention leads to a lack of productivity, in this case to the point of making contributors stop uploading any videos.


In summary, by analyzing a massive data set from \texttt{YouTube} we have shown that the productivity exhibited in crowdsourcing exhibits a strong positive dependence on attention. Conversely, a lack of attention leads to a decrease in the number of videos uploaded and the consequent drop in productivity, which in many cases asymptotes to no uploads whatsoever.  Moreover, we were able to determine that uploaders compare themselves to others when having low productivity and to themselves when exceeding a personal threshold. More generally, these results show that the tragedy of the digital commons is partly overcome by making the uploading of digital content a private good paid for by attention.

\end{document}